\documentclass[rsi,reprint]{revtex4-1}
\usepackage{graphicx}% Include figure files
\usepackage{dcolumn}% Align table columns on decimal point
\usepackage{bm}% bold math
\usepackage[mathlines]{lineno}% Enable numbering of text and display math
\usepackage[utf8]{inputenc}
\usepackage[T1]{fontenc}
\usepackage{mathptmx}
\usepackage{etoolbox}
\usepackage{color}
\begin{document}

%\preprint{}

\title{Active control of a diode laser with injection locking} 



\author{Ziting Chen}
\thanks{These authors contributed equally to this work.}
\affiliation{Department of Physics, The Hong Kong University of Science and Technology, Clear Water Bay, Kowloon, Hong Kong, China}

\author{Bojeong Seo}
\thanks{These authors contributed equally to this work.}
\affiliation{Department of Physics, The Hong Kong University of Science and Technology, Clear Water Bay, Kowloon, Hong Kong, China}

\author{Mingchen Huang}
\affiliation{Department of Physics, The Hong Kong University of Science and Technology, Clear Water Bay, Kowloon, Hong Kong, China}

\author{Mithilesh K. Parit}
\affiliation{Department of Physics, The Hong Kong University of Science and Technology, Clear Water Bay, Kowloon, Hong Kong, China}

\author{Peng Chen}
\email{pengchen@ust.hk}
\affiliation{Department of Physics, The Hong Kong University of Science and Technology, Clear Water Bay, Kowloon, Hong Kong, China}

\author{Gyu-Boong Jo}
\email{gbjo@ust.hk}
\affiliation{Department of Physics, The Hong Kong University of Science and Technology, Clear Water Bay, Kowloon, Hong Kong, China}

%\author{Ziting Chen$^{1}$, Bojeong Seo$^{1}$, Mingchen Huang$^{1}$, Mithilesh K. Parit$^{1}$, Peng Chen$^{1,a)}$, Gyu-Boong Jo$^{1,b)}$}
%\affiliation{Department of Physics, The Hong Kong University of Science and Technology, Clear Water Bay, Kowloon, Hong Kong, China}
%\email{$^{\dagger}$ These authors contributed equally to this work. \authormark{*}pengchen@ust.hk or gbjo@ust.hk} 
%\thanks{}

\date{\today}

\begin{abstract}
	We present a simple and effective method to implement an active stabilization of a diode laser with injection locking, which requires minimal user intervenes.  The injection locked state of the diode laser is probed by a photodetector,  of which sensitivity is enhanced by a narrow laser-line filter. Taking advantage of the characteristic response of laser power to spectral modes from the narrow laser-line filter,  we demonstrate that high spectral purity and low intensity noise of the diode can be simultaneously maintained by an active feedback to the injected laser. Our method is intrinsically cost-effective, and does not require bulky devices, such as Fabry-Perot interferometers or wavemeters, to actively stabilize the diode laser. Based on successful implementation of this method in our quantum gas experiments, it is conceivable that our active stabilization will greatly simplify potential applications of injection locking of diode lasers in modularized or integrated optical systems.
\end{abstract}

\maketitle %\maketitle must follow title, authors, abstract and \pacs

\section{Introduction}
Injection locking of a diode laser is essential in a majority of fields, such as laser spectroscopy~\cite{wieman1991using, eng1980tunable}, laser cooling and trapping~\cite{schafer2015optical, komori2003injection, hosoya2015injection, shimada2013simplified, pagett2016injection, schkolnik2020generating}, optical communication~\cite{goldberg1983microwave, pang2020hacking, paraiso2019modulator} and high-precision metrology~\cite{diddams2004standards, takamoto2005optical, hinkley2013atomic, liu2015selection}. It allows phase and frequency locking of a "slave" laser in reference to a "master" laser by an optical link~\cite{hadley1986injection}. Among the advantages of using injection locking of the diode are its cost-effectiveness, flexibility and efficiency. For example, it offers a simple solution to provide a clean laser source without a special design of laser diodes or controllers. The spectral mode of injection locked slave laser is nearly dependent on the master laser. Most of single mode diode lasers, including some multimode diode lasers can be used for injection locking. This feature allows us to harness injection locking for achieving a relatively high power single-frequency diode laser source, without using an optical grating that causes significant optical loss. Beside those advantages, the injection locking scheme can also be exploited as active optical filters in quantum optics~\cite{tistomo2011laser}.

In spite of its broad applications, it remains a challenge that an injection locked laser requires mechanically and thermally isolated environment otherwise necessitates an unpredictable re-locking by an user. A major obstacle is the stability of injection locking which is hard to actively maintain. Environmental variations and current drifts of laser controller can frequently cause the slave laser out of lock. It is not convenient to manually re-lock the slave laser, especially for injection locking of multiple laser sources. In order to actively stabilize the injection locking, it is proposed to use a Fabry-Perot interferometer (FPI) to monitor a locking state of the slave laser~\cite{saxberg2016active}. The FPI operates in a scanning mode to detect the spectral mode of the slave laser and thereby determines its locking state. As the FPI itself has frequency drifts, the scanning mode is necessary for proper detection in their method. However, a FPI is sensitive to optical beam alignment and the scanning mode significantly sacrifices the detection bandwidth, making it unfavorable for general-purpose applications.

\begin{figure}[b]
	\centering{\includegraphics[width=\linewidth]{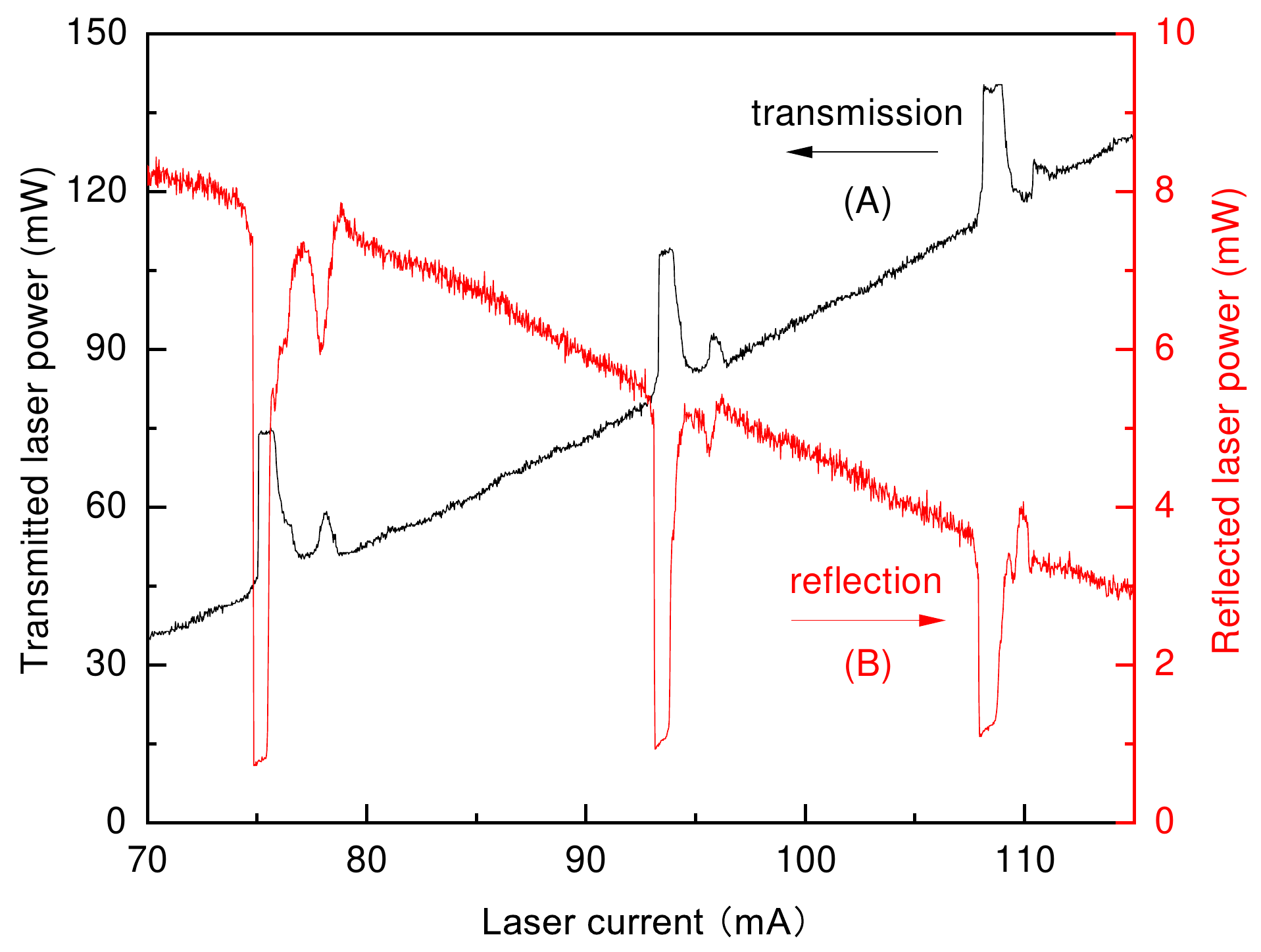}}
	\caption{Measured laser power as a function of diode current.  (A). Transmitted laser power from a narrow laser-line filter shows a series of spikes (black solid line). At those spikes, the injected slave laser operates exactly on the same frequency of the master laser. (B). Reflected laser power from the narrow laser-line filter exhibits a series of dips (red solid line), which will be used for active locking. }
	\label{fig1}
\end{figure}

\begin{figure*}
	\centering{\includegraphics[width=0.9\linewidth]{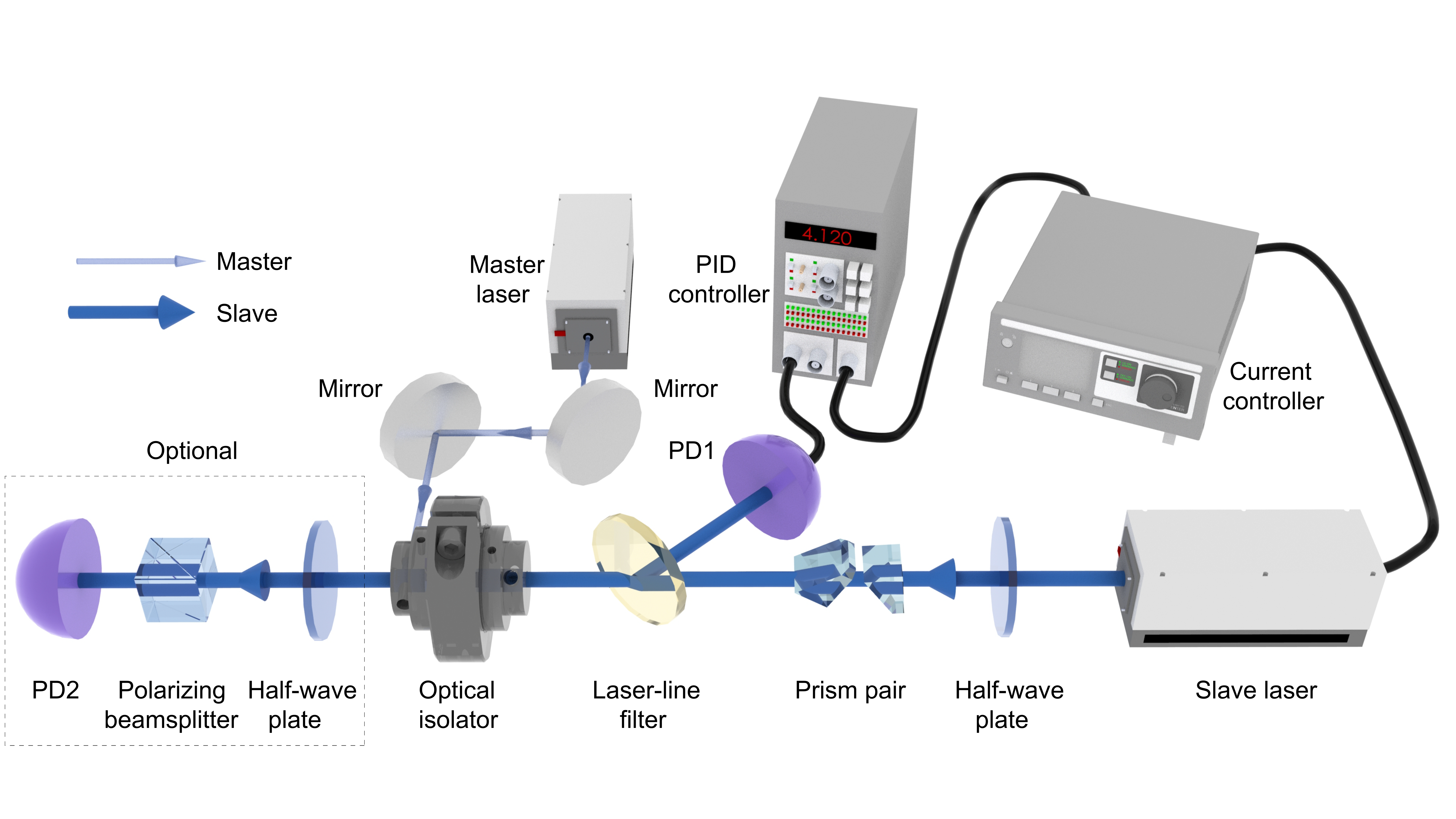}}
	\caption{Schematics of the active stabilization scheme. A narrow laser-line filter is inserted into the beam path of injection locking, from which the reflected beam is collected by a photodetector (PD1) and sent to a PID controller. The PID controller generates feedback signal to regulate the slave laser current. Optionally, the transmitted power, which will be useful for applications, can be measured with another photodetector (PD2).}
	\label{fig2}
\end{figure*}

In this work, we propose a simple way to effectively implement active control of laser injection locking without requiring additional bulky devices. We employ a photodetector and a narrow laser-line filter to probe the spectral mode of slave laser, which can be easily integrated into the beam path. The spectral mode of slave laser is monitored through the reflected signal from the narrow-laser line filter, which is  neither sensitive to beam alignment nor bandwidth limited.  With a simple feedback to the slave laser current, we demonstrate that high spectral mode purity and low intensity noise of the slave laser can be simultaneously obtained. We also show that the slave laser can be locked even when the master laser frequency is dynamically scanned over GHz range. We have successfully applied this method in laser cooling and trapping of erbium atoms towards degenerate dipolar quantum gases~\cite{lu2011strongly, aikawa2012bose}.  Our method paves the way towards general applications of injection locking. A promising scenario is coherent addition of multiple laser diodes for light power scaling,  where injection locking plays a fundamental role~\cite{schimmel2017coherent}.

\section{Active control scheme for injection locking of a diode laser}
\subsection{Working principle} 
A diode laser with light injection assumes power spikes versus laser current above threshold. These power spikes happen when the internal cavity of diode laser is near resonant to the frequency of injection light. The current threshold for lasing decreases as coherent addition of light power from the external injection beam. Therefore, the total output laser power increases in the injection locking area. As the diode laser frequency is tunable with current, the injection locking is usually done by adjusting the current.

Fig.~\ref{fig1}(A) shows a typical output power-current curve from an injected laser diode (transmitted from a laser-line filter), where output power is enhanced only at certain currents. Those power spikes reflect that the internal cavity mode of the diode is resonant to the injection light. The shape of spikes are asymmetrical due to the nonlinear response of lasing to the external injection light in contrast to the passive cavities,  where the transmission follows a Lorenz curve.  As the power spikes are correlated to injection locking, we can use them to monitor locking state of the diode laser. This can be done with a simple photodetector to monitor the laser power. The detection sensitivity can be further enhanced by a narrow laser-line filter~\cite{schkolnik2020generating}, which typically has a transmission bandwidth of 2-3 nm and high optical density (OD$>$5) for side bands rejection. The central transmission rate of laser-line filters can be also very high ($>$90\%).  The photodetector measures the reflection beam of the diode laser from the laser-line filter, where a local minimum power indicates an injection locking point, conditioning that injection light frequency is within the transmission bandwidth. With this detection geometry, the DC offset of the photodetector can be largely removed, resulting in an enhanced signal-noise ratio compared to monitoring the transmitted laser power, as shown in Fig.~\ref{fig1}(B).

\subsection{Schematics of the active stabilization scheme} 
A schematic diagram of the active injection locking of the slave laser diode is shown in Fig. 2. An external cavity diode laser (ECDL, Toptica DL Pro 401 nm), a master laser, provides 0.7~mW light for injecting a slave laser (Nichia, NDV4316, 400-410 nm) through an optical isolator (Thorlabs IO-3D-405-PBS). The optical isolator maintains bi-directional isolation for master laser and slave laser beams to avoid optical crosstalk.
A narrow laser-line filter (Semrock LL01-405-25) is inserted in the beam path before the optical isolator. This filter is slightly rotated in order to shift the centeral frequency from 405~nm to 401~nm. The rotation angle is determined by minimizing the reflective power of seeding light. A photodetector (Thorlabs PDA36A2) is used to probe the light reflected from the laser-line filter. A digital proportional-integral-derivative (PID) module (Toptica Digilock 110) regulates the slave laser current based on the photodetector output.

\begin{figure}%[!htb]
	\centering{\includegraphics[width=\linewidth]{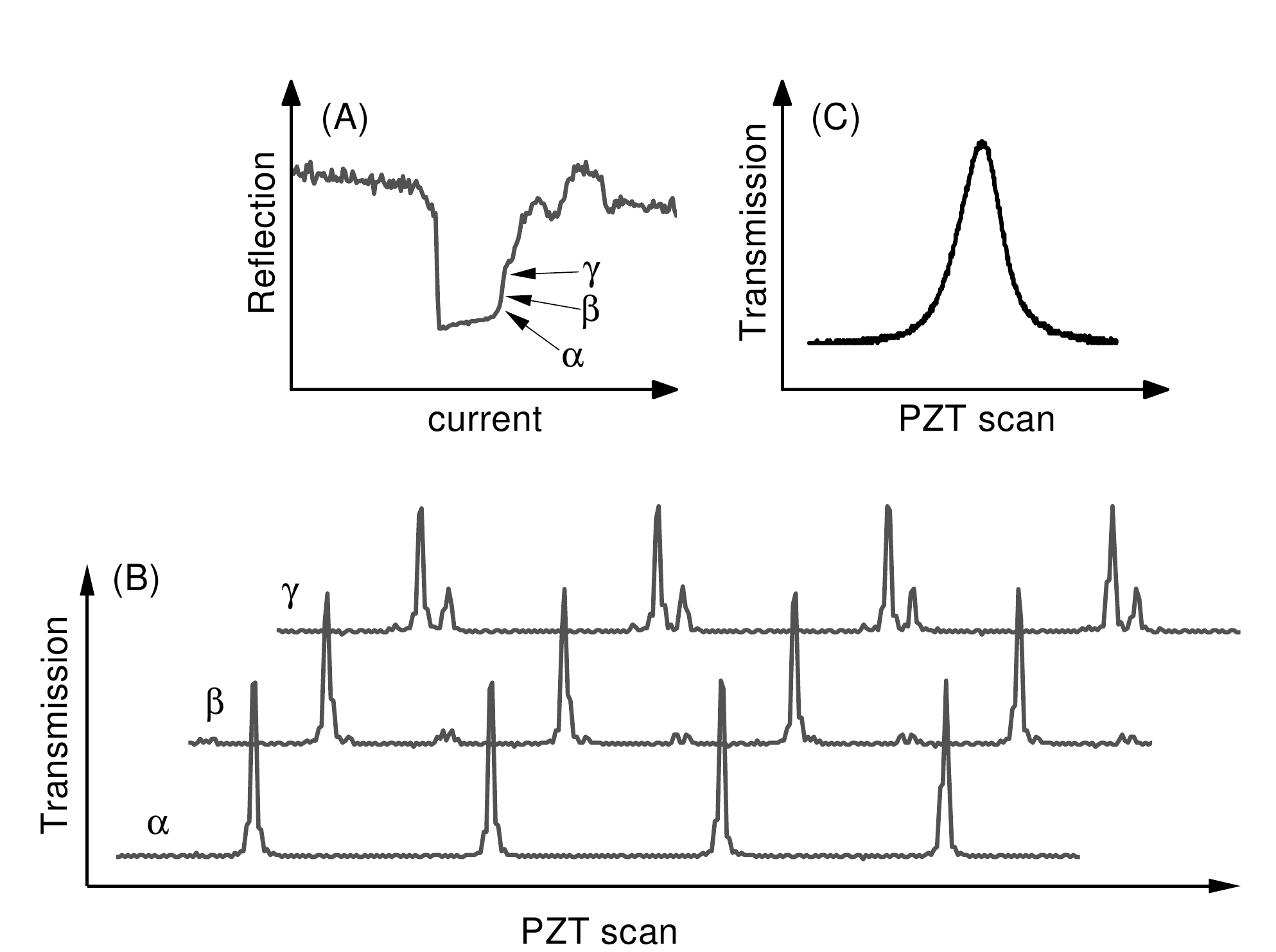}}
	\caption{Emission spectrum of the injected slave laser under active stabilization (A) We examine spectral purity of the slave laser at different lock positions. Three positions $\alpha,\beta,\gamma$ are chosen in the dip of the reflected laser power from the narrow laser-line filter. (B). FPI transmission signals of the slave laser at locking positions $\alpha,\beta,\gamma$ (C). Zoom-in profile of a single peak in curve $\alpha$ of (B).}
	\label{fig3}
\end{figure}

As described above, a local minimum of the photodetector output corresponds to an injection locking point of the slave laser. One may find an optimal injection locking current by searching the desired output power of the slave laser. However, we find it difficult to control the injected slave laser through the  peak-locking because of the asymmetric current-power response including minimum plateaus. By noting that the locking state is less sensitive to the current on the right side of the power dip of reflected laser beam, we take advantage of the side-of-fringe locking for active control of the slave laser current.  (see Fig.~\ref{fig1} and Fig.~\ref{fig3} (A)).  In contrast to the previous work where a special algorithm is required to search for optimal current for injection locking~\cite{saxberg2016active}, our scheme does not require constant searching mode set by an algorithm. Consequently, the active locking scheme does not increase the intensity noise of the slave laser without limiting the feedback bandwidth.

%This is different to the reported PID method, where a special algorithm is required to track current positions for injection locking. In their algorithm \cite{Active}, because the PID circuit does not know which direction is right for next step, the PID algorithm is always kept in a searching mode, although the searching step can be very small. The searching mode has a risk to increase intensity noise of the slave laser, and it limits the feedback bandwidth too. As a comparison, side-of-fringe locking is one of the most common PID methods.

\subsection{Spectral purity with the active control} 
\begin{figure}%[!htb]
	\centering{\includegraphics[width=\linewidth]{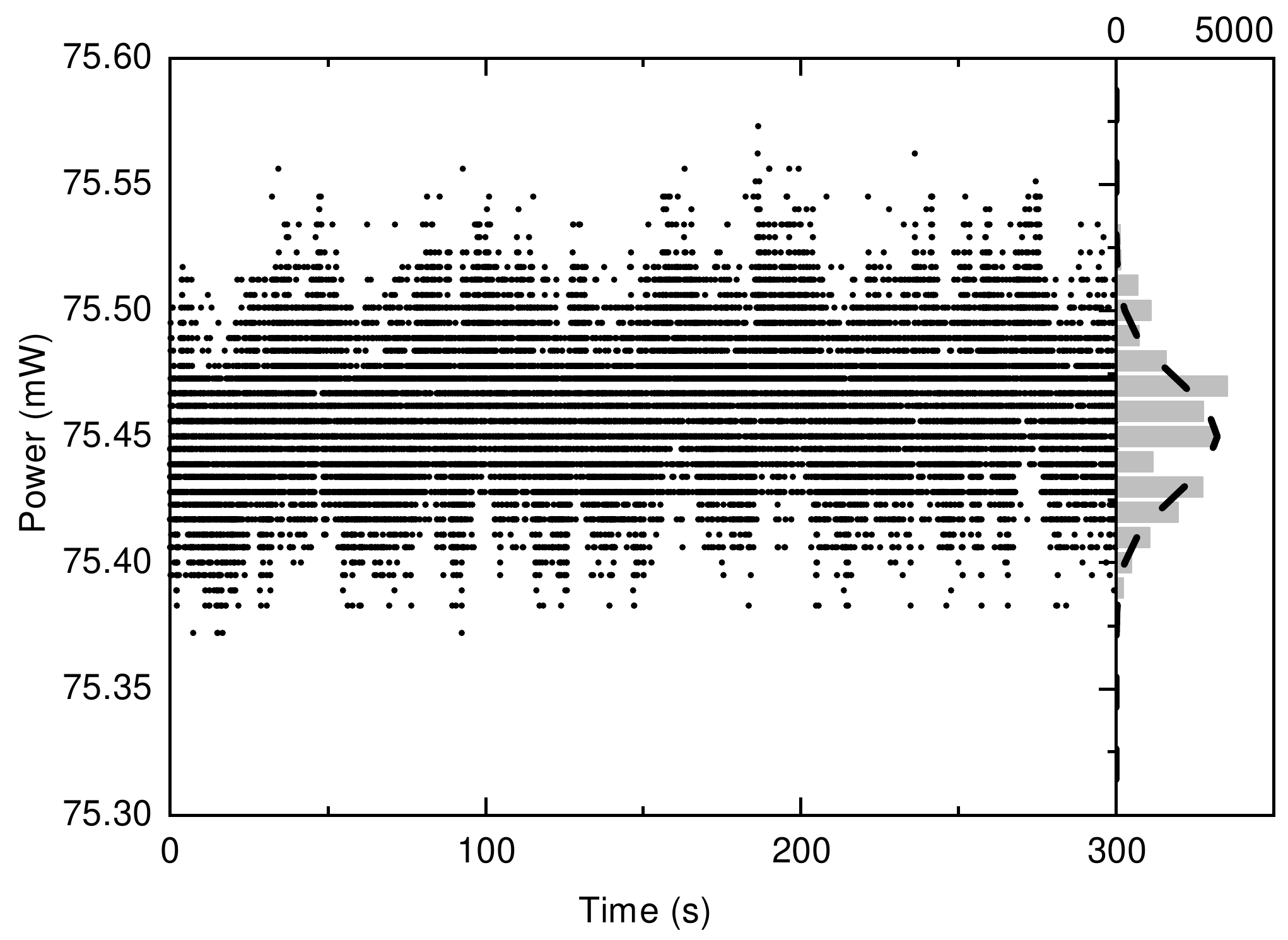}}
	\caption{Power stability of the injected slave laser. We monitor the emission power of the slave laser for 6~minutes at the position of optional part (see Fig.~2 ). Histogram of laser power is shown in the right panel. }
	\label{fig4}
\end{figure}

In Fig.~\ref{fig3}, we examine how the side-of-fringe locking affects the spectral purity for different locking position. With a slave laser operating at 25$^{\circ}$C and 109~mA,  we monitor the spectral mode of the injected laser with a Fabry-Perot interferometry (FPI, Thorlabs SA210-3B, not shown in Fig. 2). In a typical reflection signal (Fig.~\ref{fig3} (a)), we stabilize the injection locking on the right side of the dip using the side-of-fringe locking scheme. At position $\alpha$, closest to the minimum of the reflection signal, the FPI measurement does not present any spurious modes, which suggests that the slave laser frequency is faithfully locked to the master laser. At position $\beta$, spurious modes slightly show up and become more severe at position $\gamma$, which deviates most from the dip minimum among three positions. These results demonstrate that injection locking can be maintained with a proper lock position using the side-of-fringe locking. Our active locking scheme maintains injection locking at the laser power slightly higher than the minimum of the reflected laser power (e.g. the position $\alpha$), and stabilizes the power fluctuation of the slave laser. To evaluate the stability of slave laser power under the condition of injection locking with active control, we measure the slave laser power transmitted from the laser-line filter.  The ratio of the root-mean-square (rms) value of laser power fluctuations to the mean value of laser power is approximately 0.06\% in the course of 6 minutes measurement.  Referring to the slave laser power stability  without the active control, which is about 0.08 to 0.1\% (not shown in Fig.~\ref{fig4}), we find that our active control scheme can not only stabilize the injection locking but also reduce the laser intensity noise.

\subsection{Master laser frequency scan, locking bandwidth} 
\begin{figure}%[!htb]
	\centering{\includegraphics[width=\linewidth]{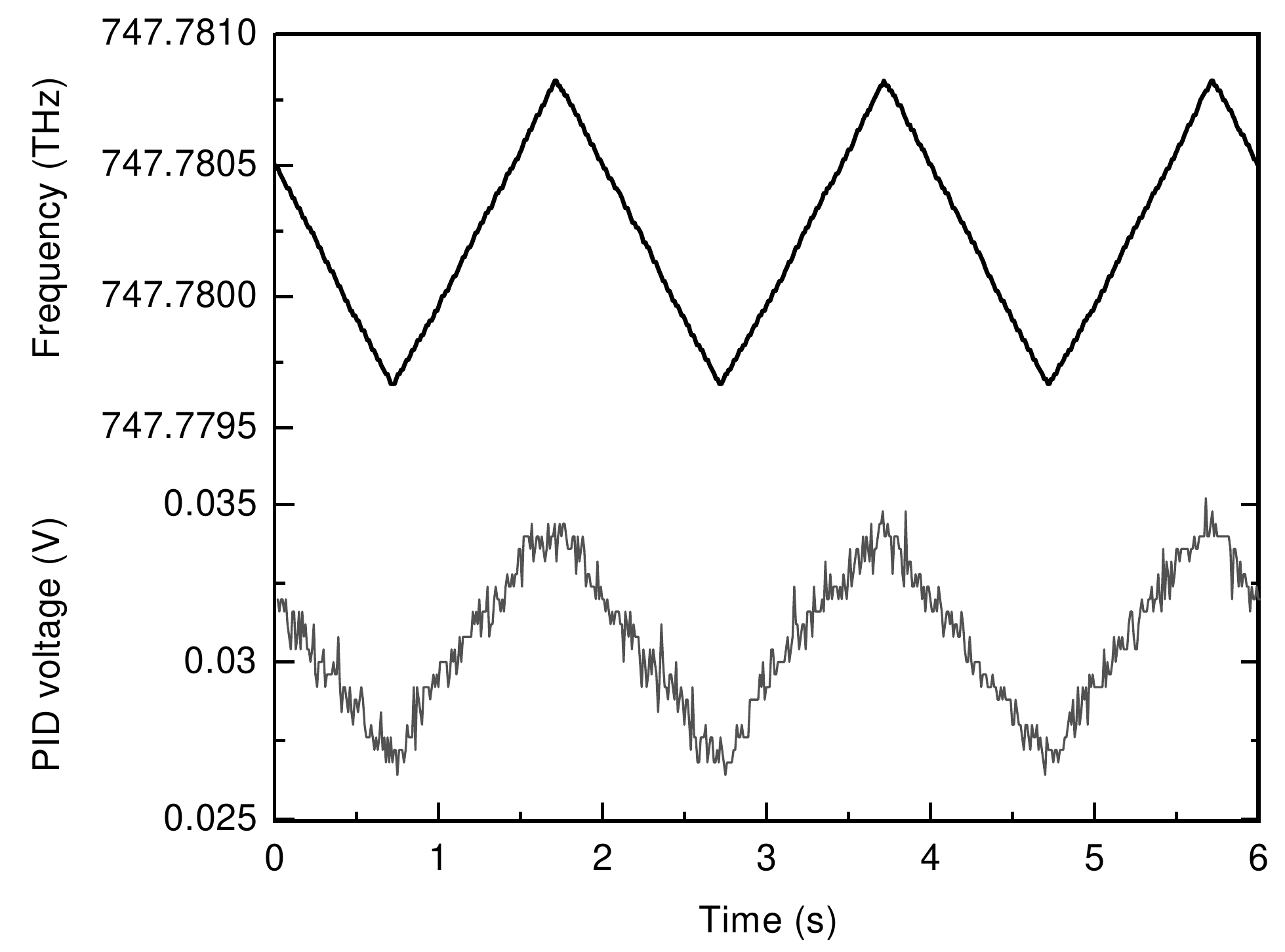}}
	\caption{Active feedback of injection locking with scanning of master laser frequency. The wavelength of slave laser (top) is measured with scanning of master laser frequency over 1.4~GHz. The corresponding PID controller output voltage (bottom) shows that the active feedback compensates the shift of injection locking point.}
	\label{fig5}
\end{figure}

Another advantage of this method is that the frequency of slave laser can be dynamically tuned, following the master laser frequency. Fig.~\ref{fig5} shows the wavelength (top) of the slave laser and the corresponding PID output voltage (bottom) from the feedback controller when the master laser linearly scans over 1~GHz. The change of the injection locking point induced by the master laser is readily compensated by the active feedback in contrast to the passive injection locking. Our results suggest that the demonstrated active feedback method is useful for not only the frequency stabilization, but also dynamically tracking the injection locking states. This feature would be critical in applications where the slave laser frequency is also required to vary, for example, coherent light detection and ranging (LIDAR) for range detection \cite{zhang2019laser}. The feedback bandwidth of our locking scheme is mainly determined by the current controller (Thorlabs, LDC202C, 100 kHz) of the slave laser. We have evaluated the feedback bandwidth by modulating the current of the master laser.

\section{Application in laser cooling of cold atoms}
Recent developments in cooling non-alkali atoms for quantum simulation~\cite{he2019recent} and computing~\cite{daley2011quantum} require stable high-power laser sources at the wavelength of 400-500~nm. In early experiments, those wavelengths are obtained by frequency doubling infrared lasers because of the limited power of the blue laser diode. The active locking scheme demonstrated in this work will provide a cost-effective way for cooling non-alkali cold atoms, such as ytterbium~\cite{song2016cost} and erbium~\cite{seo2020efficient} without using such a frequency doubling setup. 

The scheme is already being used in the daily production of Bose-Einstein condensates of erbium atoms ($^{168}$Er) in our lab~\cite{seo2020efficient}.  We employ two slave lasers to provide sufficient power for slowing erbium atoms from an effusive oven.  One slave laser delivers its full power to the major Zeeman slower, up to 30 mW after fiber coupling. The other slave laser provides similar power, solely used for the second stage Zeeman slower~\cite{seo2020efficient}. They are injection locked from the same master laser with the active locking scheme demonstrated in this work. The master laser frequency is externally locked to the erbium 401 nm transition ($4f^{12}6s^2\ ^3H_6$ $\longrightarrow$ $4f^{12}(^3H_6) 6s6p(^1P_1)$). We expect our method to be applicable to general diode lasers for injection locking, especially in ultraviolet wavelength, where high power light source is rare and typically achieved by frequency doubling of infrared light~\cite{adams1992tunable, pizzocaro2014efficient}. 

\section{Conclusion}
In conclusion, we have demonstrated a novel scheme for active control of laser injection locking. We have shown that the active stabilization of injection locking can be easily implemented with simple optical elements but without a special PID algorithm being required.  The spectral purity and power noise of the slave laser are evaluated. We demonstrate that the frequency of slave laser can be dynamically tuned following the seeding light. All these features ensure that our active control method offers a new possibility of harnessing injection locking of diode lasers in modularized or integrated optical system. Furthermore, we envision operating a diode laser with automatic injection locking, which sets the stage for achieving an auto-lock laser system.

\section*{Acknowledgments}
P.C. acknowledges the support by Shanghai Natural Science Foundation (Grant No. 18ZR1443800) and Innovation Promotion Association of the Chinese Academy of Sciences. G.-B.J. acknowledges supports from the RGC, the Croucher Foundation and the Hari Harilela foundation (project 16305317, 16304918, 16306119,16302420, C6005-17G and N-HKUST601/17).

\section*{Data availability}
The data that support the findings of this study are available from the corresponding author upon reasonable request.

%\section*{References}
\bibliography{Reference_injection}

\begin{thebibliography}{28}%
\makeatletter
\providecommand \@ifxundefined [1]{%
 \@ifx{#1\undefined}
}%
\providecommand \@ifnum [1]{%
 \ifnum #1\expandafter \@firstoftwo
 \else \expandafter \@secondoftwo
 \fi
}%
\providecommand \@ifx [1]{%
 \ifx #1\expandafter \@firstoftwo
 \else \expandafter \@secondoftwo
 \fi
}%
\providecommand \natexlab [1]{#1}%
\providecommand \enquote  [1]{``#1''}%
\providecommand \bibnamefont  [1]{#1}%
\providecommand \bibfnamefont [1]{#1}%
\providecommand \citenamefont [1]{#1}%
\providecommand \href@noop [0]{\@secondoftwo}%
\providecommand \href [0]{\begingroup \@sanitize@url \@href}%
\providecommand \@href[1]{\@@startlink{#1}\@@href}%
\providecommand \@@href[1]{\endgroup#1\@@endlink}%
\providecommand \@sanitize@url [0]{\catcode `\\12\catcode `\$12\catcode
  `\&12\catcode `\#12\catcode `\^12\catcode `\_12\catcode `\%12\relax}%
\providecommand \@@startlink[1]{}%
\providecommand \@@endlink[0]{}%
\providecommand \url  [0]{\begingroup\@sanitize@url \@url }%
\providecommand \@url [1]{\endgroup\@href {#1}{\urlprefix }}%
\providecommand \urlprefix  [0]{URL }%
\providecommand \Eprint [0]{\href }%
\providecommand \doibase [0]{http://dx.doi.org/}%
\providecommand \selectlanguage [0]{\@gobble}%
\providecommand \bibinfo  [0]{\@secondoftwo}%
\providecommand \bibfield  [0]{\@secondoftwo}%
\providecommand \translation [1]{[#1]}%
\providecommand \BibitemOpen [0]{}%
\providecommand \bibitemStop [0]{}%
\providecommand \bibitemNoStop [0]{.\EOS\space}%
\providecommand \EOS [0]{\spacefactor3000\relax}%
\providecommand \BibitemShut  [1]{\csname bibitem#1\endcsname}%
\let\auto@bib@innerbib\@empty
%</preamble>
\bibitem [{\citenamefont {Wieman}\ and\ \citenamefont
  {Hollberg}(1991)}]{wieman1991using}%
  \BibitemOpen
  \bibfield  {author} {\bibinfo {author} {\bibfnamefont {C.~E.}\ \bibnamefont
  {Wieman}}\ and\ \bibinfo {author} {\bibfnamefont {L.}~\bibnamefont
  {Hollberg}},\ }\href@noop {} {\bibfield  {journal} {\bibinfo  {journal} {Rev.
  Sci. Instrum.}\ }\textbf {\bibinfo {volume} {62}},\ \bibinfo {pages} {1}
  (\bibinfo {year} {1991})}\BibitemShut {NoStop}%
\bibitem [{\citenamefont {Eng}\ \emph {et~al.}(1980)\citenamefont {Eng},
  \citenamefont {Butler},\ and\ \citenamefont {Linden}}]{eng1980tunable}%
  \BibitemOpen
  \bibfield  {author} {\bibinfo {author} {\bibfnamefont {R.}~\bibnamefont
  {Eng}}, \bibinfo {author} {\bibfnamefont {J.}~\bibnamefont {Butler}}, \ and\
  \bibinfo {author} {\bibfnamefont {K.}~\bibnamefont {Linden}},\ }\href@noop {}
  {\bibfield  {journal} {\bibinfo  {journal} {Opt. Eng.}\ }\textbf {\bibinfo
  {volume} {19}},\ \bibinfo {pages} {196945} (\bibinfo {year}
  {1980})}\BibitemShut {NoStop}%
\bibitem [{\citenamefont {Sch{\"a}fer}\ \emph {et~al.}(2015)\citenamefont
  {Sch{\"a}fer}, \citenamefont {Ballance}, \citenamefont {Tock},\ and\
  \citenamefont {Lucas}}]{schafer2015optical}%
  \BibitemOpen
  \bibfield  {author} {\bibinfo {author} {\bibfnamefont {V.}~\bibnamefont
  {Sch{\"a}fer}}, \bibinfo {author} {\bibfnamefont {C.}~\bibnamefont
  {Ballance}}, \bibinfo {author} {\bibfnamefont {C.}~\bibnamefont {Tock}}, \
  and\ \bibinfo {author} {\bibfnamefont {D.}~\bibnamefont {Lucas}},\
  }\href@noop {} {\bibfield  {journal} {\bibinfo  {journal} {Opt. Lett.}\
  }\textbf {\bibinfo {volume} {40}},\ \bibinfo {pages} {4265} (\bibinfo {year}
  {2015})}\BibitemShut {NoStop}%
\bibitem [{\citenamefont {Komori}\ \emph {et~al.}(2003)\citenamefont {Komori},
  \citenamefont {Takasu}, \citenamefont {Kumakura}, \citenamefont {Takahashi},\
  and\ \citenamefont {Yabuzaki}}]{komori2003injection}%
  \BibitemOpen
  \bibfield  {author} {\bibinfo {author} {\bibfnamefont {K.}~\bibnamefont
  {Komori}}, \bibinfo {author} {\bibfnamefont {Y.}~\bibnamefont {Takasu}},
  \bibinfo {author} {\bibfnamefont {M.}~\bibnamefont {Kumakura}}, \bibinfo
  {author} {\bibfnamefont {Y.}~\bibnamefont {Takahashi}}, \ and\ \bibinfo
  {author} {\bibfnamefont {T.}~\bibnamefont {Yabuzaki}},\ }\href@noop {}
  {\bibfield  {journal} {\bibinfo  {journal} {Jpn. journal applied physics}\
  }\textbf {\bibinfo {volume} {42}},\ \bibinfo {pages} {5059} (\bibinfo {year}
  {2003})}\BibitemShut {NoStop}%
\bibitem [{\citenamefont {Hosoya}\ \emph {et~al.}(2015)\citenamefont {Hosoya},
  \citenamefont {Miranda}, \citenamefont {Inoue},\ and\ \citenamefont
  {Kozuma}}]{hosoya2015injection}%
  \BibitemOpen
  \bibfield  {author} {\bibinfo {author} {\bibfnamefont {T.}~\bibnamefont
  {Hosoya}}, \bibinfo {author} {\bibfnamefont {M.}~\bibnamefont {Miranda}},
  \bibinfo {author} {\bibfnamefont {R.}~\bibnamefont {Inoue}}, \ and\ \bibinfo
  {author} {\bibfnamefont {M.}~\bibnamefont {Kozuma}},\ }\href@noop {}
  {\bibfield  {journal} {\bibinfo  {journal} {Rev. Sci. Instrum.}\ }\textbf
  {\bibinfo {volume} {86}},\ \bibinfo {pages} {073110} (\bibinfo {year}
  {2015})}\BibitemShut {NoStop}%
\bibitem [{\citenamefont {Shimada}\ \emph {et~al.}(2013)\citenamefont
  {Shimada}, \citenamefont {Chida}, \citenamefont {Ohtsubo}, \citenamefont
  {Aoki}, \citenamefont {Takeuchi}, \citenamefont {Kuga},\ and\ \citenamefont
  {Torii}}]{shimada2013simplified}%
  \BibitemOpen
  \bibfield  {author} {\bibinfo {author} {\bibfnamefont {Y.}~\bibnamefont
  {Shimada}}, \bibinfo {author} {\bibfnamefont {Y.}~\bibnamefont {Chida}},
  \bibinfo {author} {\bibfnamefont {N.}~\bibnamefont {Ohtsubo}}, \bibinfo
  {author} {\bibfnamefont {T.}~\bibnamefont {Aoki}}, \bibinfo {author}
  {\bibfnamefont {M.}~\bibnamefont {Takeuchi}}, \bibinfo {author}
  {\bibfnamefont {T.}~\bibnamefont {Kuga}}, \ and\ \bibinfo {author}
  {\bibfnamefont {Y.}~\bibnamefont {Torii}},\ }\href@noop {} {\bibfield
  {journal} {\bibinfo  {journal} {Rev. Sci. Instrum.}\ }\textbf {\bibinfo
  {volume} {84}},\ \bibinfo {pages} {063101} (\bibinfo {year}
  {2013})}\BibitemShut {NoStop}%
\bibitem [{\citenamefont {Pagett}\ \emph {et~al.}(2016)\citenamefont {Pagett},
  \citenamefont {Moriya}, \citenamefont {Celistrino~Teixeira}, \citenamefont
  {Shiozaki}, \citenamefont {Hemmerling},\ and\ \citenamefont
  {Courteille}}]{pagett2016injection}%
  \BibitemOpen
  \bibfield  {author} {\bibinfo {author} {\bibfnamefont {C.}~\bibnamefont
  {Pagett}}, \bibinfo {author} {\bibfnamefont {P.}~\bibnamefont {Moriya}},
  \bibinfo {author} {\bibfnamefont {R.}~\bibnamefont {Celistrino~Teixeira}},
  \bibinfo {author} {\bibfnamefont {R.}~\bibnamefont {Shiozaki}}, \bibinfo
  {author} {\bibfnamefont {M.}~\bibnamefont {Hemmerling}}, \ and\ \bibinfo
  {author} {\bibfnamefont {P.~W.}\ \bibnamefont {Courteille}},\ }\href@noop {}
  {\bibfield  {journal} {\bibinfo  {journal} {Rev. Sci. Instrum.}\ }\textbf
  {\bibinfo {volume} {87}},\ \bibinfo {pages} {053105} (\bibinfo {year}
  {2016})}\BibitemShut {NoStop}%
\bibitem [{\citenamefont {Schkolnik}\ \emph {et~al.}(2020)\citenamefont
  {Schkolnik}, \citenamefont {Williams},\ and\ \citenamefont
  {Yu}}]{schkolnik2020generating}%
  \BibitemOpen
  \bibfield  {author} {\bibinfo {author} {\bibfnamefont {V.}~\bibnamefont
  {Schkolnik}}, \bibinfo {author} {\bibfnamefont {J.~R.}\ \bibnamefont
  {Williams}}, \ and\ \bibinfo {author} {\bibfnamefont {N.}~\bibnamefont
  {Yu}},\ }\href@noop {} {\bibfield  {journal} {\bibinfo  {journal} {arXiv
  preprint arXiv:2004.11732}\ } (\bibinfo {year} {2020})}\BibitemShut {NoStop}%
\bibitem [{\citenamefont {Goldberg}\ \emph {et~al.}(1983)\citenamefont
  {Goldberg}, \citenamefont {Taylor}, \citenamefont {Weller},\ and\
  \citenamefont {Bloom}}]{goldberg1983microwave}%
  \BibitemOpen
  \bibfield  {author} {\bibinfo {author} {\bibfnamefont {L.}~\bibnamefont
  {Goldberg}}, \bibinfo {author} {\bibfnamefont {H.}~\bibnamefont {Taylor}},
  \bibinfo {author} {\bibfnamefont {J.}~\bibnamefont {Weller}}, \ and\ \bibinfo
  {author} {\bibfnamefont {D.}~\bibnamefont {Bloom}},\ }\href@noop {}
  {\bibfield  {journal} {\bibinfo  {journal} {Electron. Lett.}\ }\textbf
  {\bibinfo {volume} {19}},\ \bibinfo {pages} {491} (\bibinfo {year}
  {1983})}\BibitemShut {NoStop}%
\bibitem [{\citenamefont {Pang}\ \emph {et~al.}(2020)\citenamefont {Pang},
  \citenamefont {Yang}, \citenamefont {Zhang}, \citenamefont {Dou},
  \citenamefont {Li}, \citenamefont {Gao},\ and\ \citenamefont
  {Jin}}]{pang2020hacking}%
  \BibitemOpen
  \bibfield  {author} {\bibinfo {author} {\bibfnamefont {X.-L.}\ \bibnamefont
  {Pang}}, \bibinfo {author} {\bibfnamefont {A.-L.}\ \bibnamefont {Yang}},
  \bibinfo {author} {\bibfnamefont {C.-N.}\ \bibnamefont {Zhang}}, \bibinfo
  {author} {\bibfnamefont {J.-P.}\ \bibnamefont {Dou}}, \bibinfo {author}
  {\bibfnamefont {H.}~\bibnamefont {Li}}, \bibinfo {author} {\bibfnamefont
  {J.}~\bibnamefont {Gao}}, \ and\ \bibinfo {author} {\bibfnamefont {X.-M.}\
  \bibnamefont {Jin}},\ }\href@noop {} {\bibfield  {journal} {\bibinfo
  {journal} {Phys. Rev. Appl.}\ }\textbf {\bibinfo {volume} {13}},\ \bibinfo
  {pages} {034008} (\bibinfo {year} {2020})}\BibitemShut {NoStop}%
\bibitem [{\citenamefont {Para{\"\i}so}\ \emph {et~al.}(2019)\citenamefont
  {Para{\"\i}so}, \citenamefont {De~Marco}, \citenamefont {Roger},
  \citenamefont {Marangon}, \citenamefont {Dynes}, \citenamefont {Lucamarini},
  \citenamefont {Yuan},\ and\ \citenamefont {Shields}}]{paraiso2019modulator}%
  \BibitemOpen
  \bibfield  {author} {\bibinfo {author} {\bibfnamefont {T.~K.}\ \bibnamefont
  {Para{\"\i}so}}, \bibinfo {author} {\bibfnamefont {I.}~\bibnamefont
  {De~Marco}}, \bibinfo {author} {\bibfnamefont {T.}~\bibnamefont {Roger}},
  \bibinfo {author} {\bibfnamefont {D.~G.}\ \bibnamefont {Marangon}}, \bibinfo
  {author} {\bibfnamefont {J.~F.}\ \bibnamefont {Dynes}}, \bibinfo {author}
  {\bibfnamefont {M.}~\bibnamefont {Lucamarini}}, \bibinfo {author}
  {\bibfnamefont {Z.}~\bibnamefont {Yuan}}, \ and\ \bibinfo {author}
  {\bibfnamefont {A.~J.}\ \bibnamefont {Shields}},\ }\href@noop {} {\bibfield
  {journal} {\bibinfo  {journal} {NPJ Quantum Inf.}\ }\textbf {\bibinfo
  {volume} {5}},\ \bibinfo {pages} {1} (\bibinfo {year} {2019})}\BibitemShut
  {NoStop}%
\bibitem [{\citenamefont {Diddams}\ \emph {et~al.}(2004)\citenamefont
  {Diddams}, \citenamefont {Bergquist}, \citenamefont {Jefferts},\ and\
  \citenamefont {Oates}}]{diddams2004standards}%
  \BibitemOpen
  \bibfield  {author} {\bibinfo {author} {\bibfnamefont {S.~A.}\ \bibnamefont
  {Diddams}}, \bibinfo {author} {\bibfnamefont {J.~C.}\ \bibnamefont
  {Bergquist}}, \bibinfo {author} {\bibfnamefont {S.~R.}\ \bibnamefont
  {Jefferts}}, \ and\ \bibinfo {author} {\bibfnamefont {C.~W.}\ \bibnamefont
  {Oates}},\ }\href@noop {} {\bibfield  {journal} {\bibinfo  {journal}
  {Science}\ }\textbf {\bibinfo {volume} {306}},\ \bibinfo {pages} {1318}
  (\bibinfo {year} {2004})}\BibitemShut {NoStop}%
\bibitem [{\citenamefont {Takamoto}\ \emph {et~al.}(2005)\citenamefont
  {Takamoto}, \citenamefont {Hong}, \citenamefont {Higashi},\ and\
  \citenamefont {Katori}}]{takamoto2005optical}%
  \BibitemOpen
  \bibfield  {author} {\bibinfo {author} {\bibfnamefont {M.}~\bibnamefont
  {Takamoto}}, \bibinfo {author} {\bibfnamefont {F.-L.}\ \bibnamefont {Hong}},
  \bibinfo {author} {\bibfnamefont {R.}~\bibnamefont {Higashi}}, \ and\
  \bibinfo {author} {\bibfnamefont {H.}~\bibnamefont {Katori}},\ }\href@noop {}
  {\bibfield  {journal} {\bibinfo  {journal} {Nature}\ }\textbf {\bibinfo
  {volume} {435}},\ \bibinfo {pages} {321} (\bibinfo {year}
  {2005})}\BibitemShut {NoStop}%
\bibitem [{\citenamefont {Hinkley}\ \emph {et~al.}(2013)\citenamefont
  {Hinkley}, \citenamefont {Sherman}, \citenamefont {Phillips}, \citenamefont
  {Schioppo}, \citenamefont {Lemke}, \citenamefont {Beloy}, \citenamefont
  {Pizzocaro}, \citenamefont {Oates},\ and\ \citenamefont
  {Ludlow}}]{hinkley2013atomic}%
  \BibitemOpen
  \bibfield  {author} {\bibinfo {author} {\bibfnamefont {N.}~\bibnamefont
  {Hinkley}}, \bibinfo {author} {\bibfnamefont {J.~A.}\ \bibnamefont
  {Sherman}}, \bibinfo {author} {\bibfnamefont {N.~B.}\ \bibnamefont
  {Phillips}}, \bibinfo {author} {\bibfnamefont {M.}~\bibnamefont {Schioppo}},
  \bibinfo {author} {\bibfnamefont {N.~D.}\ \bibnamefont {Lemke}}, \bibinfo
  {author} {\bibfnamefont {K.}~\bibnamefont {Beloy}}, \bibinfo {author}
  {\bibfnamefont {M.}~\bibnamefont {Pizzocaro}}, \bibinfo {author}
  {\bibfnamefont {C.~W.}\ \bibnamefont {Oates}}, \ and\ \bibinfo {author}
  {\bibfnamefont {A.~D.}\ \bibnamefont {Ludlow}},\ }\href@noop {} {\bibfield
  {journal} {\bibinfo  {journal} {Science}\ }\textbf {\bibinfo {volume}
  {341}},\ \bibinfo {pages} {1215} (\bibinfo {year} {2013})}\BibitemShut
  {NoStop}%
\bibitem [{\citenamefont {Liu}\ \emph {et~al.}(2015)\citenamefont {Liu},
  \citenamefont {Yin}, \citenamefont {Kong}, \citenamefont {Xu}, \citenamefont
  {Zhang},\ and\ \citenamefont {Chang}}]{liu2015selection}%
  \BibitemOpen
  \bibfield  {author} {\bibinfo {author} {\bibfnamefont {H.}~\bibnamefont
  {Liu}}, \bibinfo {author} {\bibfnamefont {M.}~\bibnamefont {Yin}}, \bibinfo
  {author} {\bibfnamefont {D.}~\bibnamefont {Kong}}, \bibinfo {author}
  {\bibfnamefont {Q.}~\bibnamefont {Xu}}, \bibinfo {author} {\bibfnamefont
  {S.}~\bibnamefont {Zhang}}, \ and\ \bibinfo {author} {\bibfnamefont
  {H.}~\bibnamefont {Chang}},\ }\href@noop {} {\bibfield  {journal} {\bibinfo
  {journal} {Appl. Phys. Lett.}\ }\textbf {\bibinfo {volume} {107}},\ \bibinfo
  {pages} {151104} (\bibinfo {year} {2015})}\BibitemShut {NoStop}%
\bibitem [{\citenamefont {Hadley}(1986)}]{hadley1986injection}%
  \BibitemOpen
  \bibfield  {author} {\bibinfo {author} {\bibfnamefont {G.}~\bibnamefont
  {Hadley}},\ }\href@noop {} {\bibfield  {journal} {\bibinfo  {journal} {IEEE
  J. Quantum Electron.}\ }\textbf {\bibinfo {volume} {22}},\ \bibinfo {pages}
  {419} (\bibinfo {year} {1986})}\BibitemShut {NoStop}%
\bibitem [{\citenamefont {Tistomo}\ and\ \citenamefont
  {Gee}(2011)}]{tistomo2011laser}%
  \BibitemOpen
  \bibfield  {author} {\bibinfo {author} {\bibfnamefont {A.~S.}\ \bibnamefont
  {Tistomo}}\ and\ \bibinfo {author} {\bibfnamefont {S.}~\bibnamefont {Gee}},\
  }\href@noop {} {\bibfield  {journal} {\bibinfo  {journal} {Opt. Express}\
  }\textbf {\bibinfo {volume} {19}},\ \bibinfo {pages} {1081} (\bibinfo {year}
  {2011})}\BibitemShut {NoStop}%
\bibitem [{\citenamefont {Saxberg}\ \emph {et~al.}(2016)\citenamefont
  {Saxberg}, \citenamefont {Plotkin-Swing},\ and\ \citenamefont
  {Gupta}}]{saxberg2016active}%
  \BibitemOpen
  \bibfield  {author} {\bibinfo {author} {\bibfnamefont {B.}~\bibnamefont
  {Saxberg}}, \bibinfo {author} {\bibfnamefont {B.}~\bibnamefont
  {Plotkin-Swing}}, \ and\ \bibinfo {author} {\bibfnamefont {S.}~\bibnamefont
  {Gupta}},\ }\href@noop {} {\bibfield  {journal} {\bibinfo  {journal} {Rev.
  Sci. Instrum.}\ }\textbf {\bibinfo {volume} {87}},\ \bibinfo {pages} {063109}
  (\bibinfo {year} {2016})}\BibitemShut {NoStop}%
\bibitem [{\citenamefont {Lu}\ \emph {et~al.}(2011)\citenamefont {Lu},
  \citenamefont {Burdick}, \citenamefont {Youn},\ and\ \citenamefont
  {Lev}}]{lu2011strongly}%
  \BibitemOpen
  \bibfield  {author} {\bibinfo {author} {\bibfnamefont {M.}~\bibnamefont
  {Lu}}, \bibinfo {author} {\bibfnamefont {N.~Q.}\ \bibnamefont {Burdick}},
  \bibinfo {author} {\bibfnamefont {S.~H.}\ \bibnamefont {Youn}}, \ and\
  \bibinfo {author} {\bibfnamefont {B.~L.}\ \bibnamefont {Lev}},\ }\href@noop
  {} {\bibfield  {journal} {\bibinfo  {journal} {Phys. Rev. Lett.}\ }\textbf
  {\bibinfo {volume} {107}},\ \bibinfo {pages} {190401} (\bibinfo {year}
  {2011})}\BibitemShut {NoStop}%
\bibitem [{\citenamefont {Aikawa}\ \emph {et~al.}(2012)\citenamefont {Aikawa},
  \citenamefont {Frisch}, \citenamefont {Mark}, \citenamefont {Baier},
  \citenamefont {Rietzler}, \citenamefont {Grimm},\ and\ \citenamefont
  {Ferlaino}}]{aikawa2012bose}%
  \BibitemOpen
  \bibfield  {author} {\bibinfo {author} {\bibfnamefont {K.}~\bibnamefont
  {Aikawa}}, \bibinfo {author} {\bibfnamefont {A.}~\bibnamefont {Frisch}},
  \bibinfo {author} {\bibfnamefont {M.}~\bibnamefont {Mark}}, \bibinfo {author}
  {\bibfnamefont {S.}~\bibnamefont {Baier}}, \bibinfo {author} {\bibfnamefont
  {A.}~\bibnamefont {Rietzler}}, \bibinfo {author} {\bibfnamefont
  {R.}~\bibnamefont {Grimm}}, \ and\ \bibinfo {author} {\bibfnamefont
  {F.}~\bibnamefont {Ferlaino}},\ }\href@noop {} {\bibfield  {journal}
  {\bibinfo  {journal} {Phys. Rev. Lett.}\ }\textbf {\bibinfo {volume} {108}},\
  \bibinfo {pages} {210401} (\bibinfo {year} {2012})}\BibitemShut {NoStop}%
\bibitem [{\citenamefont {Schimmel}\ \emph {et~al.}(2017)\citenamefont
  {Schimmel}, \citenamefont {Janicot}, \citenamefont {Hanna}, \citenamefont
  {Decker}, \citenamefont {Crump}, \citenamefont {Erbert}, \citenamefont
  {Witte}, \citenamefont {Traub}, \citenamefont {Georges},\ and\ \citenamefont
  {Lucas-Leclin}}]{schimmel2017coherent}%
  \BibitemOpen
  \bibfield  {author} {\bibinfo {author} {\bibfnamefont {G.}~\bibnamefont
  {Schimmel}}, \bibinfo {author} {\bibfnamefont {S.}~\bibnamefont {Janicot}},
  \bibinfo {author} {\bibfnamefont {M.}~\bibnamefont {Hanna}}, \bibinfo
  {author} {\bibfnamefont {J.}~\bibnamefont {Decker}}, \bibinfo {author}
  {\bibfnamefont {P.}~\bibnamefont {Crump}}, \bibinfo {author} {\bibfnamefont
  {G.}~\bibnamefont {Erbert}}, \bibinfo {author} {\bibfnamefont
  {U.}~\bibnamefont {Witte}}, \bibinfo {author} {\bibfnamefont
  {M.}~\bibnamefont {Traub}}, \bibinfo {author} {\bibfnamefont
  {P.}~\bibnamefont {Georges}}, \ and\ \bibinfo {author} {\bibfnamefont
  {G.}~\bibnamefont {Lucas-Leclin}},\ }in\ \href@noop {} {\emph {\bibinfo
  {booktitle} {High-Power Diode Laser Technology XV}}},\ Vol.\ \bibinfo
  {volume} {10086}\ (\bibinfo {organization} {International Society for Optics
  and Photonics},\ \bibinfo {year} {2017})\ p.\ \bibinfo {pages}
  {100860O}\BibitemShut {NoStop}%
\bibitem [{\citenamefont {Zhang}\ \emph {et~al.}(2019)\citenamefont {Zhang},
  \citenamefont {Pouls},\ and\ \citenamefont {Wu}}]{zhang2019laser}%
  \BibitemOpen
  \bibfield  {author} {\bibinfo {author} {\bibfnamefont {X.}~\bibnamefont
  {Zhang}}, \bibinfo {author} {\bibfnamefont {J.}~\bibnamefont {Pouls}}, \ and\
  \bibinfo {author} {\bibfnamefont {M.~C.}\ \bibnamefont {Wu}},\ }\href@noop {}
  {\bibfield  {journal} {\bibinfo  {journal} {Opt. Express}\ }\textbf {\bibinfo
  {volume} {27}},\ \bibinfo {pages} {9965} (\bibinfo {year}
  {2019})}\BibitemShut {NoStop}%
\bibitem [{\citenamefont {He}\ \emph {et~al.}(2019)\citenamefont {He},
  \citenamefont {Hajiyev}, \citenamefont {Ren}, \citenamefont {Song},\ and\
  \citenamefont {Jo}}]{he2019recent}%
  \BibitemOpen
  \bibfield  {author} {\bibinfo {author} {\bibfnamefont {C.}~\bibnamefont
  {He}}, \bibinfo {author} {\bibfnamefont {E.}~\bibnamefont {Hajiyev}},
  \bibinfo {author} {\bibfnamefont {Z.}~\bibnamefont {Ren}}, \bibinfo {author}
  {\bibfnamefont {B.}~\bibnamefont {Song}}, \ and\ \bibinfo {author}
  {\bibfnamefont {G.-B.}\ \bibnamefont {Jo}},\ }\href@noop {} {\bibfield
  {journal} {\bibinfo  {journal} {J. Phys. B: At. Mol. Opt. Phys.}\ }\textbf
  {\bibinfo {volume} {52}},\ \bibinfo {pages} {102001} (\bibinfo {year}
  {2019})}\BibitemShut {NoStop}%
\bibitem [{\citenamefont {Daley}(2011)}]{daley2011quantum}%
  \BibitemOpen
  \bibfield  {author} {\bibinfo {author} {\bibfnamefont {A.~J.}\ \bibnamefont
  {Daley}},\ }\href@noop {} {\bibfield  {journal} {\bibinfo  {journal} {Quantum
  Inf. Process.}\ }\textbf {\bibinfo {volume} {10}},\ \bibinfo {pages} {865}
  (\bibinfo {year} {2011})}\BibitemShut {NoStop}%
\bibitem [{\citenamefont {Song}\ \emph {et~al.}(2016)\citenamefont {Song},
  \citenamefont {Zou}, \citenamefont {Zhang}, \citenamefont {Cho},\ and\
  \citenamefont {Jo}}]{song2016cost}%
  \BibitemOpen
  \bibfield  {author} {\bibinfo {author} {\bibfnamefont {B.}~\bibnamefont
  {Song}}, \bibinfo {author} {\bibfnamefont {Y.}~\bibnamefont {Zou}}, \bibinfo
  {author} {\bibfnamefont {S.}~\bibnamefont {Zhang}}, \bibinfo {author}
  {\bibfnamefont {C.-w.}\ \bibnamefont {Cho}}, \ and\ \bibinfo {author}
  {\bibfnamefont {G.-B.}\ \bibnamefont {Jo}},\ }\href@noop {} {\bibfield
  {journal} {\bibinfo  {journal} {Appl. Phys. B}\ }\textbf {\bibinfo {volume}
  {122}},\ \bibinfo {pages} {1} (\bibinfo {year} {2016})}\BibitemShut {NoStop}%
\bibitem [{\citenamefont {Seo}\ \emph {et~al.}(2020)\citenamefont {Seo},
  \citenamefont {Chen}, \citenamefont {Chen}, \citenamefont {Yuan},
  \citenamefont {Huang}, \citenamefont {Du},\ and\ \citenamefont
  {Jo}}]{seo2020efficient}%
  \BibitemOpen
  \bibfield  {author} {\bibinfo {author} {\bibfnamefont {B.}~\bibnamefont
  {Seo}}, \bibinfo {author} {\bibfnamefont {P.}~\bibnamefont {Chen}}, \bibinfo
  {author} {\bibfnamefont {Z.}~\bibnamefont {Chen}}, \bibinfo {author}
  {\bibfnamefont {W.}~\bibnamefont {Yuan}}, \bibinfo {author} {\bibfnamefont
  {M.}~\bibnamefont {Huang}}, \bibinfo {author} {\bibfnamefont
  {S.}~\bibnamefont {Du}}, \ and\ \bibinfo {author} {\bibfnamefont {G.-B.}\
  \bibnamefont {Jo}},\ }\href@noop {} {\bibfield  {journal} {\bibinfo
  {journal} {Phys. Rev. A}\ }\textbf {\bibinfo {volume} {102}},\ \bibinfo
  {pages} {013319} (\bibinfo {year} {2020})}\BibitemShut {NoStop}%
\bibitem [{\citenamefont {Adams}\ and\ \citenamefont
  {Ferguson}(1992)}]{adams1992tunable}%
  \BibitemOpen
  \bibfield  {author} {\bibinfo {author} {\bibfnamefont {C.}~\bibnamefont
  {Adams}}\ and\ \bibinfo {author} {\bibfnamefont {A.}~\bibnamefont
  {Ferguson}},\ }\href@noop {} {\bibfield  {journal} {\bibinfo  {journal} {Opt.
  Commun.}\ }\textbf {\bibinfo {volume} {90}},\ \bibinfo {pages} {89} (\bibinfo
  {year} {1992})}\BibitemShut {NoStop}%
\bibitem [{\citenamefont {Pizzocaro}\ \emph {et~al.}(2014)\citenamefont
  {Pizzocaro}, \citenamefont {Calonico}, \citenamefont {Pastor}, \citenamefont
  {Catani}, \citenamefont {Costanzo}, \citenamefont {Levi},\ and\ \citenamefont
  {Lorini}}]{pizzocaro2014efficient}%
  \BibitemOpen
  \bibfield  {author} {\bibinfo {author} {\bibfnamefont {M.}~\bibnamefont
  {Pizzocaro}}, \bibinfo {author} {\bibfnamefont {D.}~\bibnamefont {Calonico}},
  \bibinfo {author} {\bibfnamefont {P.~C.}\ \bibnamefont {Pastor}}, \bibinfo
  {author} {\bibfnamefont {J.}~\bibnamefont {Catani}}, \bibinfo {author}
  {\bibfnamefont {G.~A.}\ \bibnamefont {Costanzo}}, \bibinfo {author}
  {\bibfnamefont {F.}~\bibnamefont {Levi}}, \ and\ \bibinfo {author}
  {\bibfnamefont {L.}~\bibnamefont {Lorini}},\ }\href@noop {} {\bibfield
  {journal} {\bibinfo  {journal} {Appl. optics}\ }\textbf {\bibinfo {volume}
  {53}},\ \bibinfo {pages} {3388} (\bibinfo {year} {2014})}\BibitemShut
  {NoStop}%
\end{thebibliography}%

\end{document}